\documentstyle[12pt,epsf]{article}
\newcommand{\be}{\begin{equation}}
\newcommand{\ee}{\end{equation}}
\newcommand{\al}{\mbox{$\alpha$}}

\newcommand{\bi}[1]{\bibitem{#1}}
\newcommand{\fr}[2]{\frac{#1}{#2}}

\newcommand{\Ima}{\mbox{Im}}

\begin{document}
\begin{titlepage}
\rightline{\vbox{\halign{&#\hfil\cr
&UQAM-PHE-96/10\cr
&CUMQ/HEP 94\cr
&\today\cr}}}
\vspace{0.5in}
\begin{center}
{\bf  
$\mu\rightarrow e\gamma$ DECAY IN THE LEFT-RIGHT SUPERSYMMETRIC MODEL}
\\
\medskip
\vskip0.5in

\normalsize {{\bf G. Couture}$^{\rm a}$,
{\bf M. Frank}$^{\rm b}$, {\bf H. K\"{o}nig}$^{\rm a,b}$
and {\bf M. Pospelov}$^{\rm a,c}$}
\smallskip
\medskip

{ \sl $^a$
Depart\'{e}ment de Physique, Universit\'{e} du Qu\'{e}bec a Montr\'{e}al\\ 
C.P. 8888, Succ. Centre Ville, Montreal, Quebec, Canada, H3C 3P8\\
$^b$Department of Physics, Concordia University, 
1455 De Maisonneuve Blvd. W.\\ Montreal, Quebec, Canada, H3G 1M8 \\
$^c$Budker Institute of Nuclear Physics, Novosibirsk, 630090, Russia} 
\smallskip
\end{center}
\vskip1.0in

\noindent{\large\bf Abstract}
\smallskip

We calculate the rate of the decay $\mu\rightarrow e\gamma$ and the electric dipole moment of the electron
in the left-right supersymmetric model when the breaking of parity occurs at 
a considerably large scale. 
The low-energy flavor violation in the model originates either from the nonvanishing 
remnants of the left-right symmetry in the slepton mass matrix or from the 
direct flavor changing lepton-slepton-neutralino interaction. The result is 
found to be large for the masses of the supersymmetric particles not far from 
the electroweak scale and already accessible at the current experimental 
accuracy. It also provides nontrivial constraints of the lepton mixing in the model.  

\end{titlepage}

\baselineskip=20pt

\newpage
\pagenumbering{arabic}
\section{\bf Introduction}
\label{intro}

The quest for a supersymmetric grand unified theory is plagued by lack of direct
 signals which would distinguish such a theory from supersymmetry in general.
Supersymmetry, in particular the Minimal Supersymmetric Model (MSSM)
~\cite{habnil}, can be
probed experimentally through the  high energy production of superpartners.
 However, the MSSM, while filling in some of the theoretical gaps of the 
Standard Model, fails to explain other phenomena such as the weak mixing angle, 
the small mass (or masslessness) of the known neutrinos, the origin of CP 
violation, to mention a few. Extended gauge structures such as grand unified 
theories, introduced to provide an elegant framework for unification of 
forces~\cite{ellis}, would connect 
the standard model with more fundamental structures such as superstrings,
and also resolve the puzzles of the electroweak theory.

Phenomenologically, grand unified theories would predict either relationships 
between otherwise independent parameters of the standard model, 
or new interactions (i.e. 
interactions forbidden or highly suppressed in the standard model).

Among supersymmetric grand unified theories SO(10)~\cite{barbieri1} and 
SU(5)~\cite{barbieri2} have received significant attention. It is known that the 
rates of processes with flavor number violation, such as $\mu\rightarrow 
e\gamma$, $\mu \rightarrow 3e$ and $\mu-e$ conversion on nuclei are 
significantly enhanced in comparison with the pure MSSM case and are just one 
order of magnitude below the current experimental limit.
 In this article we  shall study a model, the left-right 
symmetric extension of the MSSM, based on $SU(2)_{L}\times SU(2)_{R}\times 
U(1)_{B-L}$ \cite{frank1,rasin,huitu}. We show that this model shares some 
interesting features of the unified models providing the nonvanishing remnants 
of the left-right symmetry in the slepton mass matrix even if the right-handed 
scale is far from the electroweak one. Its attraction is twofold: first, the 
Left-Right Supersymmetric 
Model $(LRSUSY)$ is an extension of the Minimal 
Supersymmetric Standard Model  based on left-right symmetry, and second, it could  be viewed as a low-energy 
realization of certain $SUSY-GUTs$, such as $SO(10)$. However, the left-right symmetry, being much less 
restrictive assumption than unification itself, does not 
relate the mixing angles in lepton and quark sectors. In particular, evaluating the 
$\mu\rightarrow e\gamma$ decay we show that this model can provide both smaller 
and {\em larger} rates than in unified theories depending mainly on the size of 
the mixing in the lepton sector. 

 Since the
 measurement of $\mu \rightarrow e \gamma$ has a very stringent bound
$B(\mu\rightarrow e\gamma) < 4.9\times 10^{-11}$
~\cite{particle data},
we are able to obtain values close 
to the experimental bound and restrict some of 
the parameters in the theory.

Our paper is organized as follows:
in section~\ref{lrsusy} we give a brief 
description of the model, followed by the 
analysis of different flavor-violating mechanisms in section 3. The amplitude of 
the decay $\mu \rightarrow e \gamma $ is analyzed in 
section 4. 
Our conclusions 
are reached in section~\ref{concl}. 

\section{\bf The Left-Right Supersymmetric Model}
\label{lrsusy}

The $LRSUSY$ model, based on $SU(2)_{L}\times SU(2)_{R}\times U(1)_{B-L}$,
has matter 
doublets for both left- and right- handed fermions and the corresponding left- 
and right-handed scalar partners (sleptons and squarks)~\cite{frank3}.
In the gauge sector, 
corresponding to $SU(2)_{L}$ and $SU(2)_{R}$, there are triplet
gauge bosons $(W^{+,-},W^{0})_{L}$, $(W^{+,-},W^{0})_{R}$ and a singlet gauge
boson $V$ corresponding to $U(1)_{B-L}$, together with their superpartners. 
The Higgs sector of this model 
consists of two Higgs bi-doublets, $\Phi_{u}(\frac{1}{2},\frac{1}{2},0)$ and 
$\Phi_{d}(\frac{1}{2},\frac{1}{2},0)$, which are required to give masses to 
both the up and down quarks. In addition, the spontaneous symmetry breaking of
the group 
$SU(2)_{R}\times U(1)_{B-L}$ to the hypercharge symmetry group $U(1)_{Y}$ is 
accomplished by introducing the Higgs triplet fields $\Delta_{L}(1,0,2)$ 
and $\Delta_{R}(0,1,2)$. The choice of the triplets (versus four doublets)
is preferred because with this choice a large Majorana mass can be generated
for the right-handed neutrino and a small one for the left-handed neutrino~
\cite{mohapatra}.
In addition to the triplets $\Delta_{L,R}$, the model must contain two 
additional triplets $\delta_{L}(1,0,-2)$ and $\delta_{R}(0,1,-2)$ , with 
quantum number $B-L= -2$ to insure cancellation of the anomalies that would 
otherwise occur in the fermionic sector.

As in the standard model, in order to preserve $U(1)_{EM}$ gauge 
invariance,  only the neutral Higgs fields acquire non-zero vacuum 
expectation values $(VEV's)$. These values are:
\begin{eqnarray}
\langle \Delta_{L} \rangle = \left(\begin{array}{cc}
0&0\\v_{L}&0
\end{array}\right),
\langle \Delta_{R} \rangle = \left (\begin{array}{cc}
0&0\\v_{R}&0
\end{array}\right)~\rm{and}~
\langle \Phi \rangle = \left (\begin{array}{cc}
\kappa&0\\0&\kappa' e^{i\omega}
\end{array}\right).
\nonumber
\end{eqnarray}
$\langle \Phi \rangle$ causes the mixing of $W_{L}$ and $W_{R}$ bosons with 
$CP$-violating 
phase $\omega$. In order to simplify, we will take the $VEV's$ of the Higgs 
fields as: $\langle \Delta_{L} \rangle = 0$ and 
\begin{eqnarray}
\langle \Delta_{R} \rangle = \left (\begin{array}{cc}
0&0\\v_{R}&0
\end{array}\right),
\langle \Phi_{u} \rangle = \left (\begin{array}{cc}
\kappa_{u}&0\\0&0
\end{array}\right)~\rm{and}~
\langle \Phi_{d} \rangle = \left (\begin{array}{cc}
0&0\\0&\kappa_{d}
\end{array}\right).
\nonumber
\end{eqnarray}
Choosing $v_{L} =\kappa' =0$ satisfies the more loosely phenomenologically required hierarchy
$v_{R}~\gg~max(\kappa,\kappa')~\gg~v_{L}$ and also the required cancellation 
of flavor-changing neutral currents. The Higgs fields acquire non-zero $VEV's$ 
to break both parity and $SU(2)_{R}$.
In the first stage of breaking the right-handed gauge bosons $W_{R}$ and $Z_{R}$
aquire masses proportional to $v_{R}$ and become much heavier than the usual
(left-handed) neutral gauge bosons $W_{L}$ and $Z_{L}$, which pick up masses 
proportional to $\kappa_{u}$ and $\kappa_{d}$ at the second stage of breaking.
~\cite{frank1}

The supersymmetric sector of the model, 
while preserving left-right symmetry, has four singly-charged
charginos ( corresponding to $\tilde\lambda_{L}, 
\tilde\lambda_{R}, \tilde\phi_{u}$, and
$\tilde\phi_{d}$), in addition to $\tilde\Delta_{L}^-$ , 
$\tilde\Delta_{R}^-$ , $\tilde\delta_{L}^-$ and $\tilde\delta_{R}^-$.
The model also has eleven neutralinos, corresponding to $\tilde\lambda_{Z}$, 
$\tilde\lambda_{Z\prime}$,
$\tilde\lambda_{V}$ ,  $\tilde\phi_{1u}^0$ ,$\tilde\phi_{2u}^0$ ,
$\tilde\phi_{1d}^0$ , $\tilde\phi_{2d}^0$, $\tilde\Delta_{L}^0$,
$\tilde\Delta_{R}^0$ $\tilde\delta_{L}^0$, and
$\tilde\delta_{R}^0$.

In the scalar matter sector, the $LRSUSY$ contains two left-handed and two 
right-handed scalar fermions as partners of the ordinary leptons and quarks, 
which themselves
come in left- and right-handed doublets. In general the left- and right-handed  
scalar leptons will mix together. Some of the effects of these mixings, such as 
enhancement of the anomalous magnetic moment of the muon, have been discussed 
elsewhere~\cite{frank1}. 

\section{Sources of flavor violation in $LRSUSY$}

The interaction of fermions with scalar (Higgs) fields has the following form:
\begin{eqnarray}
{\cal L}_Y=Y_u\bar{Q}_L \Phi_u Q_R + Y_d \bar{Q}_L \Phi_d Q_R\,
+Y_\nu\bar{L}_L  \Phi_u L_R + Y_e \bar{L}_L \Phi_d L_R+\,H.c.;\\
{\cal L}_M=iF(L_L^TC^{-1}\tau_2\Delta_LL_L+
L_R^TC^{-1}\tau_2\Delta_RL_R) + H.c.,\nonumber
\label{eq:yukawa}
\end{eqnarray}
LR symmetry requires all $Y$-matrices to be hermitean in the generation space and  
$F$ matrix to be symmetric.

The off-diagonal entries in matrices $Y_\nu$, $Y_e$ and $F$ are responsible for
 the lepton flavor violation in the theory. 

In what follows we consider the effects related to the neutrino Yukawa couplings 
$Y_\nu$. The seesaw mechanism allows to have large Yukawa couplings and at the 
same time escape the constraints coming from the neutrino mass data if the scale 
of the right-handed physics is higher than $100$ TeV. Indeed, if 
 $Fv_R$ is large, the Yukawa couplings $Y_\nu$ could be very large and there is 
 the possibility of having $ Y_{\nu_\tau}$ of the order 1.
Therefore, the seesaw mechanism could provide large amount 
of flavor violation in the leptonic sector \cite{LRmu}. In what follows we will  concentrate 
 on the specific $LRSUSY$-related mechanisms for the lepton flavor 
violation.

In complete analogy with the quark sector \cite{kaon} the  
flavor changing transitions between charged leptons are proportional to the 
Dirac Yukawa couplings of neutrinos:
\be
{\cal L}_{FCNC}= 
\bar{E}_LV^\dagger Y^{diag}_\nu VE_R\phi_{2u}^0+h.c.,
\label{eq:FCNC}
\ee
where we use already the mass eigenstate representation. Here $V$ is the 
Kobayashi-Maskawa matrix in the lepton sector. The neutral Higgs particles 
associated with $\phi_{2u}^0$ have to be very heavy in order to suppress the $FCNC$ 
contribution to the neutral kaon mixing. Therefore, the direct contribution of 
this interaction to the lepton violating processes is negligibly small because corresponding one-loop diagram is suppressed as $m_{FCNC}^2$. However,
in supersymmetric model  
we have to consider other possible interactions of the same origin involving 
higgsinos:
\be
{\cal L'}_{FCNC}= 
\bar{E}_R\tilde{\phi}_{2u}^0V^\dagger Y^{diag}_\nu V\tilde{E}_L+
\tilde{E}^*_RV^\dagger Y^{diag}_\nu V\bar{\tilde{\phi}}_{2u}^0E_L~+h.c.
\label{eq:FCNC1}
\ee
Here we do not have strict phenomenological restrictions on the  mass of the 
corresponding higgsino and therefore the interaction (\ref{eq:FCNC1}) is very 
important for the muon conversion or decay. 

If the FCNC higgsino happens to be heavy we have to consider different sources of 
flavor violation related with the scalar lepton mass matrix. Flavor violating 
terms in the slepton mass matrix arise as a result of the renormalization group evolution 
from the $\Lambda_{GUT}$ scale and are caused by the admixture of the neutrino 
Yukawa couplings. 
Instead of calculating this matrix, which could be done only numerically, we 
adopt here the following anzatz for the charged slepton matrix with some 
elements of the left-right symmetry \cite{Posp}:
\be
(\tilde{E}_L^\dagger\;
\tilde{E}_R^\dagger)
\left(
\begin{array}{cc}
M_L^2+c_e \lambda_e^2+ c_\nu \lambda_\nu^2&{\cal A}_e\\
{\cal A}_e^\dagger &M_R^2+c_e' \lambda_e^2+ c_\nu' \lambda_\nu^2 ,
\end{array}
\right)
\left(\begin{array}{c}
\tilde{E}_L\\
\tilde{E}_R
\end{array}\right),
\label{eq:mass}
\ee
where ${\cal A}_e=A(M_e+a_e\lambda_e^2M_e+
a_\nu\lambda_\nu^2M_e+a'_\nu M_e\lambda_\nu^2)-M_e\mu\tan\beta$.\newline
The coefficients $c_\nu,\, c'_\nu,\,c_e,\, c'_e ,\, a_e,\, a_\nu,\,a'_\nu$ appear either 
at the tree level or in the one-loop renormalization from $\Lambda_{GUT}$. The 
 requirement of the L-R symmetry is:
\be
M_L=M_R,\; c_e= c_e',\; c_\nu= c_\nu',\; a_\nu= a_\nu'. 
\label{eq:LR}
\ee
As a result, the mass matrix (\ref{eq:mass}) differs from that of the MSSM where 
$c_\nu'=0$ and $a_\nu'=0$. The values of all these coefficients depend on many 
additional parameters and we simply assume here the following estimate: 
\be
c_\nu\sim c_\nu'\sim m_{susy}^2(16\pi^2)^{-1}\ln(\Lambda_{GUT}^2/M_{W_R}^2)\sim 
{\cal O}(m_{susy}^2).
\label{eq:coeff}
\ee
When the left-right symmetry is broken, the relations (\ref{eq:LR}) become 
approximate and we expect $\fr{M_L^2-M_R^2}{M^2}\sim 10^{-2}-10^{-1}$ 
\cite{rasin,Posp}

Next we consider the implications of these FCNC mechanisms in the $LRSUSY$ in 
lepton-flavor violating decay $\mu \rightarrow e \gamma$.

\section{\bf The Decay $\mu \rightarrow e \gamma$ }
\label{muega}

The amplitude of the $\mu\rightarrow e \gamma$ transition can be written in the 
form of the usual dipole-type interaction:
\be
{\cal M}_{\mu \rightarrow e \gamma} =\fr{e}{2}d\bar{\psi_\mu}\sigma^{\mu\nu}F_{\mu\nu}\psi_e
\label{eq:ampl}
\ee
It leads to the partial width 
\be
\Gamma_{\mu\rightarrow e \gamma}=\fr{\alpha}{4}d^2m^3_\mu
\ee
Comparing it with the standard decay width, $\Gamma_{\mu\rightarrow e 
\nu\bar{\nu}}=\fr{1}{192\pi^2}G_F^2m_{\mu}^5$ and using the experimental 
constraint on the branching ratio, we get the following limit on $d$ \cite{ft}:
\be
|d|<3\cdot10^{-25}~e\cdot cm
\label{eq:exp}
\label{eq:limit}
\ee

In what follows we calculate $d$ due to different sources of flavor violation. 
We note that there are two general classes of contributions to $d$. First, there are
terms proportional to the mass of muon divided by the square of supersymmetric mass 
scale, $m_\mu/M^2$. The second class of contributions are terms proportional to the mass of 
the tau lepton, $m_\tau/M^2$. We concentrate our analysis on the second class 
and calculate the corresponding $d$. The enhancement factor, $m_\tau/m_\mu$, 
associated with this subclass allows us to consider only the diagrams with the 
chirality flip on the slepton line and neglect at the moment all other 
contributions. In some sense, we assume from the very beginning that the mixing 
with the third generation is significant and of the same order as the mixing in 
the $\mu-e$ sector. 

To obtain the amplitude of $\mu\rightarrow e\gamma$ derived from the 
flavor-violating mechanisms in the previous section we have to
calculate the  one-loop diagrams of Fig. 1.

In the neutral higgsinos exchange the leading contribution is due to the 
superpartner of the field $\phi_{2u}^0$ corresponding to the $FCNC$ Higgs 
(\ref{eq:FCNC1}). It gives the amplitude:
\be
|d|=\fr{1}{16\pi^2}m_\tau\fr{(A-\mu\tan\beta)m_h}{M^4}
|V_{33}|^2|V_{32}V^*_{31}|Y_{\nu_\tau}^2 F(m_h^2/M^2),
\label{eq:I}
\ee
where
\be
F(x)=\fr{1}{2(1-x)^4}\left[1+4x-5x^2+4x\log x+2x^2\log x\right]
\ee
Let us suppose for a moment that at the $GUT$ scale we have the 
unification of quarks and leptons in the form of a close correspondence between 
Yukawa couplings and mixing angles, $V_{31}\sim V_{td}$, $V_{32}\sim V_{ts}$, 
$Y_{\nu_\tau}\sim Y_t$. Then our result leads to the following numerical 
estimate, where we take the common mass for all supersymmetric parameters, $M\sim 
m_h \sim(A-\mu\tan\beta)$, $F\simeq 1/12$:
\be
|d|\sim 24\cdot 10^{-25}\left(\fr{100\mbox{GeV}}{M}\right)^2~e\cdot cm
\ee
which is already larger than the current experimental limit. 
Alternatively, we can translate Eq. (\ref{eq:I}) into the following limit:
\be
|V_{33}|^2|V_{32}V^*_{31}|Y_{\nu_\tau}^2
\left(\fr{100\mbox{GeV}}{M}\right)^2< 5\cdot 10^{-5}.
\label{eq:limit1}
\ee

Another group of diagrams  proportional to $m_\tau$ which we have 
to consider here is that of Fig. 2 where 
the fermion line in the loop corresponds to the propagation of the bino, the 
superpartner of the $U(1)$ gauge boson. The flavor violation comes through  
mass insertions in the slepton line. These insertions are indicated as small 
bubbles. The superpartners of $SU(2)_L$ or $SU(2)_R$ are not interesting because 
they couple only to left- or right-handed fermions and cannot lead to  
$m_{\tau}$-proportional effects since these require a helicity flip. Using the estimate (\ref{eq:coeff}) and for 
$m_h\sim M$, the result reads as follows:
\be
|d|\sim\fr{\alpha}{4\pi}\fr{1}{20}m_\tau
\fr{A-\mu\tan\beta}{M^3}
|V_{33}|^2|V_{32}V^*_{31}|Y^4_{\nu_\tau} 
\label{eq:II}
\ee
It gives a different dependence of Yukawas, $Y_{\nu_\tau}^4$. The constraints 
which could be obtained from (\ref{eq:II}) are one--two orders of magnitude 
weaker in comparison with (\ref{eq:limit1}) , mainly due to the factor 
$4\pi\al\sim  0.1$. 

These two mechanisms exhaust all possible supersymmetric 
contributions to $d$ proportional to the mass of the tau lepton. As to the 
diagrams with the charged higgsinos, they are irrelevant for our analysis since 
we know that large $ v_{R}$ leads to the decoupling of the right-handed 
sneutrinos through the corresponding couplings in the supersymmetric $F$-term. 

Lastly, we obtain an estimate of the electric dipole moment 
of the electron in this model due to the complex phases in the lepton Yukawa 
couplings. The mechanism analogous to (\ref{eq:I}) leads to the following 
estimate:
\be
d_{EDM}\sim J \fr{1}{16\pi^2}\fr{1}{20}m_\tau\fr{(A-\mu\tan\beta)}{M^3} 
\fr{M_L^2-M_R^2}{M^2}Y^3_\tau Y_\mu,
\label{eq:EDM}
\ee
where $J=\Ima(V_{21}V^*_{31}V_{32}V_{22}^*)$ is the CP-odd rephasing invariant 
of the leptonic KM matrix and $\fr{M_L^2-M_R^2}{M^2}\sim 10^{-2}-10^{-1}$ is the 
relative difference between left- and right-handed slepton masses. The 
corresponding constraint on the CP-odd combination of mixing angles and Yukawa 
couplings reads as follows:
\be
J Y^3_\tau Y_\mu \left(\fr{100\mbox{GeV}}{M}\right)^2<10^{-4}-10^{-3}
\ee
which is definitely weaker than constraints coming from the muon decay if we 
believe that $Y_\mu\sim Y_c \sim 10^{-2}$.

\section{\bf Conclusions}
\label{concl}

We have evaluated the $\mu\rightarrow e\gamma$ decay rate and we have shown that the 
left-right supersymmetric model could provide large rates for the lepton flavor 
violating processes if the scale of the supersymmetric masses is not far from 
the left-handed electroweak scale. The most promising mechanism in this respect 
is related with the higgsino particle, the superpartner of $FCNC$ Higgs. The 
absence of direct phenomenological requirements for this fermion to be heavy 
makes the $\mu\rightarrow e\gamma$ decay large, even without any 
flavor-changing insertions into the slepton line. The specific numerical 
predictions are weakened, however, by the lack of knowledge of explicit values for 
the supersymmetric masses. In particular, the predictive power of the model cannot be 
improved unless one specifies 
a way to solve the $FCNC$ problem . If one believes 
that the masses of $FCNC$ higgsinos are related to the $v_R$ scale and thus  
very large, one has to use another source of flavor violation coming from 
the Yukawa dependence of slepton masses. In this case $LRSUSY$ is similar to MSSM at 
the usual electroweak scale and the right-handed scale plays the role of the 
intermediate scale  considered in Ref. \cite{deshp}.

The $m_{\tau}$-proportionality of the amplitude for this process is very similar 
to the situation which occurs in unified theories \cite{barbieri1,barbieri2}. 
The similarity is not accidental. It is based  on the presence of non-MSSM type 
slepton masses and mixings which survive in the effective low-energy theory 
even if the scale of the new physics, unification scale or $v_R$, is very high. 
In $LRSUSY$, however, the mixing angles and masses in the lepton and quark 
sectors are not related. The rate of flavor-changing decay $\mu\rightarrow e\gamma$
may be even bigger than in the unified theories. Therefore we conclude that at
the  current experimental accuracy the model 
already predicts nontrivial limits on mixing angles in the lepton sector. 

In this letter we have concentrated only on $m_{\tau}$-proportional 
contributions to the amplitude of $\mu\rightarrow e\gamma$ decay. A complete 
analysis would include the calculation of subleading $m_{\mu}$-proportional 
terms which could be important because the number of diagrams involved is very 
large. One would have to consider also the effects of the neutralino mixing, the 
differences between masses of superparticles, etc; all these are model-dependent.
 We believe, however, that the 
large degree of arbitrariness in the choice of various parameters in the model 
cannot change the main features of the muon decay amplitude considered here.

\section{\bf Acknowledgements}
We would like to thank C. Hamzaoui for useful discussions. 
This work was funded by NSERC of Canada and FCAR du Qu\'ebec
The work of M.P. is supported by NATO Science Fellowship, N.S.E.R.C., grant \#  
189 630 and Russian Foundation for Basic Research, grant \# 95-02-04436-a.

\newpage
{\bf Figure captions}

Fig. 1. One-loop contribution to the muon decay amplitude due to flavor-changing 
lepton-slepton-neutralino interaction. The cross on the tau slpeton line 
indicates left-right mixing.

Fig. 2. One-loop contribution to the muon decay amplitude due to the 
flavor-changing insertions into the slepton line.
\vspace{2cm}

\begin{figure}[hbtp]
\begin{center}
\mbox{\epsfxsize=144mm\epsffile{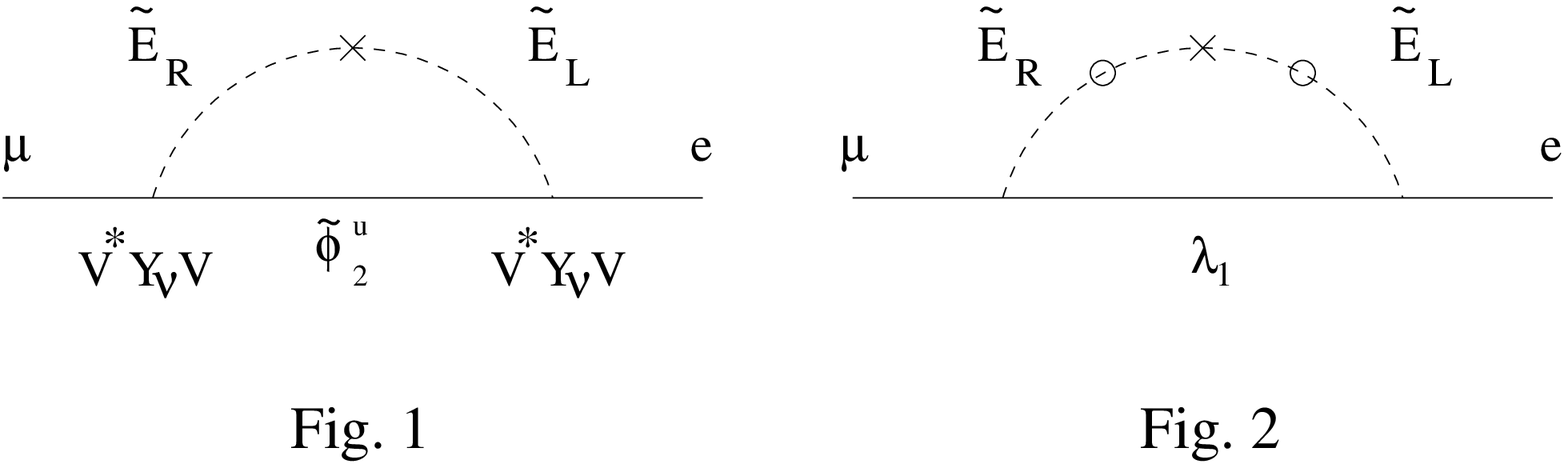}}

\end{center}
\end{figure}
\end{document}